\documentclass{amsart}
\usepackage[latin1]{inputenc}
\usepackage{amsmath,amsfonts,amssymb,amsthm}

\usepackage{times}

\DeclareMathOperator{\tr}{tr}

\title{Survey of Gravity in Non-Commutative Geometry}
\author{Nicolas Franco}

\begin{document}

\maketitle
\vspace{-0.8cm}
\begin{center}
{\small Gamasco\\
Groupe d'Applications Mathématiques aux Sciences du Cosmos\\
University of Namur FUNDP - Belgium}
\end{center}

\begin{abstract}
We present a survey of the application of Cones' Non-Commutative Geometry to gravitation. Bases of the theory and Euclidian gravity models are reviewed. Then we discuss the problem of a Lorentzian generalization of the theory and review existing attempts of solution.
\end{abstract}


\section[Motivation]{Motivation of a Non-Commutative version of Geometry}

Geometry - and especially pseudo-Riemannian geometry - has become a compulsory tool to models gravity. In particular, light deflection phenomena is a strong indication of the geometrical nature of gravity. On the other hand, Gauge theories provide us a way to construct an unified model of the three other fundamental interactions (electromagnetism, weak and strong interactions). Such theories are based on Lie algebras $\mathfrak g$, with an obviously non-abelian product operation to ensure a non-vanishing commutator \mbox{$[\cdot,\cdot] : \mathfrak g \times \mathfrak g \rightarrow \mathfrak g$}.\\

For that, we can expect that a complete mathematically coherent theory which would include both gravitation and the other fundamental interactions should be based on a mathematical background which includes at the same time gravitational aspects and non-commutative aspects. The theory of Non-Commutative Geometry - mainly developed by Alain Connes and Ali Chamseddine - is an attempt to do that, and could be interpreted as an {\it algebraization of geometry} or an {\it geometrization of algebra}.\\

In this paper, we present a short survey of the application of the theory of Non-Commutative Geometry to the problem of gravitation, in particular the difficulty to construct an unified mathematical background which includes both gravity and the Standard Model of matter.


\section[Geometry = Algebra]{Equivalence between Geometry and Algebra}

First, we review few bases of C*-Algebras theory.\\

A {\bf C*-Algebra}  is a normed complete algebra with the condition
$$\left\Vert {ab}\right\Vert \leq \left\Vert  a \right\Vert  \left\Vert  b \right\Vert $$
equipped with an involution - i.e.  a map \mbox{$a \rightarrow a^*$} such that
$$(za+b)^* = \bar z a^* + b^*\qquad(ab)^* = b^*a^*\qquad(a^*)^* = a$$
- and the following compatibility condition 
$$\left\Vert {a^*a}\right\Vert  = {\left\Vert  a \right\Vert }^2.\\$$

Let {\bf A} be a C*-Algebra, the {\bf spectrum} of {\bf A}, denoted by \mbox{$\Delta({\bf A})$}, is the set of all non zero *-morphisms $\chi : {\bf A} \rightarrow \mathbb C$, each element of the spectrum being called {\bf character}.\\

So, there is a natural map $\bigvee : {\bf A} \rightarrow C(\Delta({\bf A})) : a \leadsto \hat a$ by defining \mbox{$\hat a(\chi) = \chi(a)$}. This is the so-called {\bf Gel'fand transform}. We have then the Gel'fand-Neimark theorem :

{\it For any ({\emph{unital}}) commutative C*-algebra {\bf A}, the Gel'fand transform is an isometric isomorphism between {\bf A} and \mbox{$C_0(\Delta({\bf A}))$} ({\emph{\mbox{$C(\Delta({\bf A}))$} for the unital case}}).}\\

This leads to the fact that every locally compact ({\it or compact}) Hausdorff space can be seen as the spectrum of the ({\it unital}) commutative C*-algebra of continuous functions on this space. On the language of categories, this gives us a {\bf functor} between the category of  locally compact ({\it compact}) Hausdorff space and the category of ({\it unital}) commutative C*-algebras.


\section[Calculus]{Infinitesimal calculus}

[J. Dixmier, {\it Existence de traces non normales}, C.R. Acad. Sci. Paris, Ser. A-B (1966) 262 A1107-1108]\\

The Gel'fand-Neumark theorem guarantees that there is no {\it lost of information} by considering the algebra of continuous functions on a manifold instead of the manifold itself.\\

Now, in order to transcribe geometrical theories like General Relativity in a full algebraic framework, we should develop similar tools to those existing in the geometrical case, and in particular infinitesimal calculus.

\subsection*{Infinitesimals}

Let $\mathcal{H}$ be an Hilbert space, and $\mathcal K(\mathcal{H})$ and $\mathcal B(\mathcal{H})$ the algebras of compact and bounded operators. The compact operators will play the role of the infinitesimals. \\

We will use the very useful property of compact self-adjoint operators to have a discrete spectrum which at most one limit point in 0. The rate of decay of such sequence of eigenvalues will give us information about its order.\\

Let $T \in \mathcal K(\mathcal{H})$ and take its expansion 
$$T = \sum_{n \geq 0} \mu_n(T)\ |\phi_n>\;<\psi_n| $$
where \mbox{$\left\{{\mu_n(T), \mu_n \rightarrow 0 \text{ as } n \rightarrow \infty}\right\}$} are the characteristics values of $T$ (i.e. the ordered suite of eigenvalues of the compact self-adjoint operator \mbox{$|T| = \sqrt{T^*T}$}).\\

We define infinitesimals of order $\alpha$ by all operators \mbox{$T \in \mathcal K(\mathcal{H})$} such that
$$
\mu_n(T) = \mathcal O(n^{-\alpha}) \  \text{ as } n \rightarrow \infty.
$$

We can check the following property
$$ \mu_{n+m}(T_1 T_2) \leq \mu_n(T_1)\ \mu_m(T_2)$$
which implies the next rule :
$$ T_{1,2}\ \text{ is an infinitesimal of order }\ \alpha_{1,2}$$
$$\ \ \Longrightarrow\ \  T_1 T_2\  \text{ is an infinitesimal of order at most} \  \alpha_1 + \alpha_2.$$

\subsection*{Integrals}

We seek for an integral functional which neglects all infinitesimals of order greater than one.\\

For this fact, the usual trace for operators in \mbox{$\mathcal L^1$} 
$$ \tr(T) = \sum_{n \geq 0} \mu_n(T) $$
cannot be used because in general infinitesimals of order 1 are not necessarily in \mbox{$\mathcal L^1$}.\\

But this trace is at most logarithmically divergent
$$ \sum_{n = 0}^{N-1} \mu_n(T) \leq C \ln N $$
and it is possible to try to extract the coefficient $C$ :
$$ \lim_{N \rightarrow \infty} \frac{1}{\ln N} \sum_{n = 0}^{N-1} \mu_n(T).$$

As shown by Dixmier, this coefficient behaves like a trace.\\

The convergence of the previous limit - although bounded - is not guarantied all the time. But in such case, the limit can be replaced by a invariant scale linear form, which is called the Dixmier trace :
$$
\tr_\omega (T) = \lim\!{}_{\omega} \frac{1}{\ln N} \sum_{n = 0}^{N-1} \mu_n(T).
$$

As for infinitesimals of order higher than one we have \mbox{$n\;\mu_n(T) \rightarrow 0$}  as  \mbox{$n \rightarrow \infty$}, the corresponding Dixmier trace is convergent and converges to zero.\\

For algebras of pseudodifferential operators (of order at most $-n$) acting on sections of a vector bundle, the Dixmier trace corresponds (at least proportionally) to the Wodzicki Residue defined by :
$$Res_W(T) = \frac{1}{n(2\pi)^n} \int_{S^* M} tr_E \sigma_n(T) d\mu$$
where $\sigma_n(T) $ is the principal symbol of $T$.


\section[Spectral Triples]{Spectral Triples}

The notion of Spectral Triple is at this time probably the most important element introduced by Alain Connes in Non-Commutative Geometry.\\

A {\bf Spectral Triple} \mbox{$(\mathcal{A},\mathcal{H},D)$} is the data of
\begin{itemize}
\item an Hilbert space $\mathcal{H}$
\item an involutive algebra $\mathcal{A}$ of bounded operators on $\mathcal{H}$ 
\item an self-adjoint operator $D$ on $\mathcal{H}$ with compact resolvent such that
$$[D,a] \in \mathcal B(\mathcal{H})\quad \forall a \in \mathcal{A}\quad\text{({\it this actually corresponds to a first order condition})}$$
\end{itemize}

A spectral triple is said to be {\bf even} (odd otherwise) if there exists a \mbox{$\mathbb Z_2$}-grading $\gamma$ such that \mbox{$[\gamma,a] = 0\ \forall a\in\mathcal{A}$} and \mbox{$\left\{{\gamma,D}\right\} = 0$}.\\

A {\bf real structure of KO-dimension $n \in\mathbb Z_8$} on a spectral triple is the existence of an isometry \mbox{$J :\mathcal{H} \rightarrow \mathcal{H}$} (a sort of charge conjugation operator) such that 
$$J^2 = \epsilon\quad J D = \epsilon' D J\quad J \gamma = \epsilon'' \gamma \quad[a,JbJ^{-1}] = 0\quad[[D,a],JbJ^{-1}] = 0\ \forall a,b\in\mathcal{A}$$
with $\epsilon$, $\epsilon'$ and $\epsilon''$ with value in the set $\left\{{-1,1}\right\}$ depending of the value of $n\!\!\mod 8$.

\subsection*{Differential forms}

For any spectral triple \mbox{$(\mathcal{A},\mathcal{H},D)$} we can define an algebra of differential forms by defining a representation of the algebra of universal forms on $\mathcal{H}$ (or in the Riemannian case a representation of the algebra of usual differential forms as shown later) :
$$\pi : \Omega\mathcal{A} \rightarrow \mathcal B(\mathcal{H}) : \quad\pi(a_0\delta a_1 \cdot\cdot\cdot \delta a_p) = a_0 [D,a_1] \cdot\cdot\cdot[D,a_p]\qquad a_0,...,a_p\in\mathcal{A}$$

But this is actually not a true differential algebra since we have the problem
$$\pi(\omega) = 0\ \text{  do not imply }\ \pi(\delta \omega) = 0.$$

Such forms are called {\it junk form} and can be removed by making a quotient by the two-sided ideal $J = J_0 + \delta J_0$ where
$$J_0 = \left\{{\omega : \pi(\omega) = 0}\right\}.$$

The result is the graded differential algebra due to Connes \mbox{$\Omega_D \mathcal{A}$}.

\subsection*{Distance}

A notion of distance can be defined, given two {\it states} (positive linear functionals) on $\mathcal{A}$, $\chi$ and $\phi$, by
$$ d(\chi,\phi) = \sup_{a\in\mathcal{A}}\left\{{\left\vert {\chi(a) - \phi(a)} \right\vert: \left\Vert {[D,a]}\right\Vert  \leq 1}\right\}.$$


\section[Gravity]{Gravity in Non-Commutative Geometry}

[A. Connes, {\it Gravity coupled with matter and the foundation of noncommutative geometry}, Comm. Math. Phys. 182 (1996) 155]\\

Euclidian Gravity can be quite easily transcribed in the language of Non-Commutative Geometry. Given a compact Riemannian spin manifold $(M,g)$ of dimension 4, we construct the following real even spectral triple :
\begin{itemize}
\item $\mathcal{A} = C^\infty(M)$
\item $\mathcal{H} =  L^2(S)$
\item $D = \gamma^a e_a^\mu \left({ \partial_{\mu} + \frac{1}{4} \omega_\mu^{ab}\gamma_{ab}}\right)$
\item $\gamma = \gamma_1\gamma_2\gamma_3\gamma_4$ ({\it chirality element})
\item $J \psi = C \bar\psi \qquad \forall \psi \in\mathcal{H}$
\end{itemize}
where $S$ is the irreductible spinor bundle over $M$, \mbox{$\omega_\mu^{ab}$} the spin-connection and $C$ the charge conjugation operator.\\

We define the action of $\mathcal{A}$  on $\mathcal{H}$ simply by multiplication
$$(f\psi)(x) = f(x)\,\psi(x),\qquad \forall f\in\mathcal{A},\ \forall\psi\in\mathcal{H}$$
which implies that $[D,f]$ acts as a multiplicative operator.\\

We have then the following results :
\begin{itemize}
\item Because states correspond to characters in the abelian case, we have the distance formula :
$$ d(x,y) = \sup_{f\in\mathcal{A}}\left\{{\left\vert{f(x) - f(y)}\right\vert : \left\Vert {[D,f]}\right\Vert  \leq 1}\right\}$$
which corresponds to the usual one :
$$ d(x,y) = \inf \int_\gamma ds\qquad \gamma \text{ path from }x\text{ to }y.$$

The great interest of this new distance formula is there is no dependence to the path $\gamma$. The "non-communative" formula only depends on points $x$, $y$  and on the Dirac operator.\\

\item The graded differential algebra $\Omega_D \mathcal{A}$ (as defined before) is isomorphic to the usual differential forms $\Lambda(M)$ by the representation
$$\pi(f_0 df_1 \cdot\cdot\cdot df_p) = f_0 [D,f_1] \cdot\cdot\cdot[D,f_p].$$

\item The Riemannian measure on $M$ is given by
$$ \int_M f = c_n \tr_\omega(f|D|^{-4}),\qquad \forall f\in\mathcal{A}.$$

\item The action functionnal given by the Dixmier trace is proportional to the Hilbert-Einstein action
$$ S = \tr_\omega(D^{2}) = \tilde c_n \int_M R\ dx.$$
\end{itemize}

We have then what we can call the reconstruction theorem :\\

Consider an (even) spectral triple \mbox{$\left({\mathcal{A},\mathcal{H},D,J,(\gamma)}\right)$} whose algebra $\mathcal{A}$ is commutative. Then there exists a compact Riemannian spin manifold M (of even dimension) whose spectral triple \mbox{$\left({C^\infty(M), L^2(S),D,J,(\gamma)}\right)$} coincides with  \mbox{$\left({\mathcal{A},\mathcal{H},D,J,(\gamma)}\right)$}.


\section[Gravity + 2 points]{Gravity and the two-point space}

[A. H. Chamseddine, G. Felder and J. Fröhlich, {\it Gravity in non-commutative geometry}, Comm. Math. Phys. 155 (1993) 205]\\

The second simplest model of gravity is the product of a four dimensional manifold and a two-point space, with the Riemannian metric taken to be the same on the two copies of the manifold.\\

Space is set as the same way as a Kaluza-Klein theory :
$$ X = M \times \mathbb Z_2.$$

Then we construct the even triple
\begin{itemize}
\item $\mathcal{A} = C^\infty(M) \oplus C^\infty(M)$
\item $\mathcal{H} =  L^2(p_*S),\quad p : X \rightarrow M$
\item $D$ is an odd equivariant first order elliptic differential operator on the space \mbox{$C^\infty(p_*S)$}, such that \mbox{$[D,f]$} is a multiplicative operator
\item $\Gamma = \left(\begin{array}{cc}\gamma & 0 \\0 & -\gamma\end{array}\right)$\\
\end{itemize}

It is more convenient to interpret $\mathcal{A}$ as the subalgebra of diagonal matrices in \mbox{$M_2(\mathbb C) \otimes C^\infty(\text{Cliff}(T^*M))$}.\\

 The most general form for $D$ is
 $$ D = \left(\begin{array}{cc}\gamma^a\epsilon_a^\mu \partial_\mu & \psi + \gamma \phi \\\psi + \gamma \phi & \gamma^a\epsilon_a^\mu \partial_\mu\end{array}\right)$$

So it is possible to construct the one form corresponding to \mbox{$\alpha = a_idb^i \in \Lambda^1$} : 
 $$ \pi(\alpha) = \left(\begin{array}{cc}\gamma^a\epsilon_a^\mu \alpha_{1\mu} & (\psi + \gamma \phi)\alpha_5 \\ (\psi + \gamma \phi)\tilde\alpha_5 & \gamma^a\epsilon_a^\mu \alpha_{2\mu}\end{array}\right)$$
 $$\alpha_{j\mu} = a_{ij}\partial_\mu b^{i}_j\qquad \alpha_5 = a_{i1}(b^i_2-b^i_1)\qquad \tilde\alpha_5 = a_{i2}(b^i_2-b^i_1)$$
 
Then the {\it junk forms} :
 $$ \pi(\alpha) = 0 \ \ \Longrightarrow\ \   \pi(d \alpha) = \left(\begin{array}{cc} -\epsilon_a^\mu\epsilon_a^\nu \partial_\mu a_{i1}\partial_\nu b^i_{1} & -2\psi\gamma^a\epsilon_a^\mu a_{i1} \partial_\mu b^i_2 \\ -2\psi\gamma^a\epsilon_a^\mu a_{i2} \partial_\mu b^i_1  & -\epsilon_a^\mu\epsilon_a^\nu \partial_\mu a_{i2}\partial_\nu b^i_{2}\end{array}\right)$$
 and so the subclass of 2-forms $\pi(d\alpha)$ modulo junk forms :
 $$\pi(d \alpha) = \left(\begin{array}{cc} \gamma^a\epsilon_a^\mu \gamma^b\epsilon_b^\nu \partial_\mu \alpha_{1\nu} + 2 \phi\psi\gamma(\alpha_5 - \tilde\alpha_5) 
 & \phi\gamma^a\epsilon_a^\mu\gamma(\partial_\mu \alpha_5 + \alpha{1\mu} - \alpha_{2\mu}) \\ 
- \phi\gamma^a\epsilon_a^\mu\gamma(\partial_\mu \tilde\alpha_5 + \alpha{1\mu} - \alpha_{2\mu})
 &\gamma^a\epsilon_a^\mu \gamma^b\epsilon_b^\nu \partial_\mu \alpha_{2\nu} + 2 \phi\psi\gamma(\alpha_5 - \tilde\alpha_5)\end{array}\right)$$
 
 The whole calculation can be continued and finally leads (with $\psi = 0$) to a kind of scalar field model coupled with gravity :
 $$ S = 2 \int_M [R - 2 \partial_\mu \sigma \partial_\nu \sigma g^{\mu\nu}] \sqrt{g}d^4 x,\qquad \phi = e^{-\sigma}.$$
 

\section[Gravity + SM]{Gravity and the Standard Model}

[A. H. Chamseddine, A. Connes and M. Marcolli, {\it Gravity and the standard model with neutrino mixing}, Adv. Theor. Math. Phys. 11 (2007) 991]\\

We need to introduce two new tools.

\subsection*{Products of spectral triples}

Given two spectral triples, which we could take even and real (the first one at least must be real), \mbox{$(\mathcal{A}_1,\mathcal{H}_1,D_1,\gamma_1,(J_1))$} and  \mbox{$(\mathcal{A}_2,\mathcal{H}_2,D_2,(\gamma_2),(J_2))$}, then the product \mbox{$(\mathcal{A},\mathcal{H},D,(\gamma),(J))$} is also a (even) (real) spectral triple set by\\
\begin{itemize}
\item $\mathcal{A} = \mathcal{A}_1 \otimes \mathcal{A}_2$
\item $\mathcal{H} =  \mathcal{H}_1 \otimes \mathcal{H}_2$
\item $D = D_1 \otimes 1 + \gamma_1 \otimes D_2$
\item ($\gamma =  \gamma_1 \otimes \gamma_2$)
\item ($J  = J_1 \otimes J_2$)
\end{itemize}

\subsection*{The Spectral Action}

Gauge degrees of freedom can be interpreted as a kind of {\it inner fluctuations} in Non-Commutative Geometry.\\

These fluctuations replace the operator $D$, by
$$ D \rightarrow D_A = D + A + J A J^{-1}, \qquad A = a_i[D,b^i] \in \Omega_D^1\mathcal{A},\quad A = A^*.$$

Then the so-called spectral action is proposed
$$S =\tr_\mathcal{H}\left({ F\left({\frac{D_A^2}{\Lambda^2}}\right)}\right)$$
where $\Lambda$ is a {\it cut off} parameter and $F$ a suitable function which cut off all eigenvalues of $D_A$ larger than $\Lambda$.\\

To compute this action, one has to calculate the square of the Dirac operator with Lichnérowicz' formula \mbox{$D^2_A = \nabla^*\nabla - \mathcal E$} and the trace in the Hilbert space with the method of heat kernel expansion.\\

We have to assume the {\it spectral action principle} which says that the dynamic depends only on the spectum of the Dirac operator.  In case of gravity, this spectral invariance is however a stronger condition than the usual diffeomorphism invariance since there exist manifolds which are isospectral without being isometric.

\subsection*{The Gravitational part}

The spectral triple \mbox{$(\mathcal{A}_1,\mathcal{H}_1,D_1,\gamma_1,J_1)$} is chosen as described in the Gravity section.\\

The spectral action associated is computed by heat kernel expansion and gives
$$ S =  \frac {1}{16\pi G}\int_M (-R + 2\lambda) \sqrt{g}d^4 x  \  +  \ \frac {f_0}{10\pi^2}\int_M (\frac{11}{6}L_{GB} - 3C_{\mu\nu\kappa\epsilon}C^{\mu\nu\kappa\epsilon}) \sqrt{g}d^4x\  +\   \mathcal O(\Lambda^{-2}). $$

\subsection*{The Standard Model part}

The spectral triple \mbox{$(\mathcal{A}_2,\mathcal{H}_2,D_2,\gamma_2,J_2)$} must be chosen in order that  the group of automophisms on $\mathcal{A}$ coincides with the gauge group of the standard model \mbox{$U(1) \times SU(2) \times SU(3)$}.\\

First we define the algebra
$$\mathcal{A}_{LR} = \mathbb C \oplus \mathbb H_L\oplus \mathbb H_R \oplus M_3(\mathbb C)$$
and denote all irreductible representations by is dimension (ex: ${\bf 2_L}$ is the 2-dimmentional irredictible representation of  $\mathbb H_L$).\\

We set the bimodule 
$$\mathcal E = {\bf 2_L} \otimes {\bf 1^0} \oplus{\bf 2_R} \otimes{\bf 1^0} \oplus{\bf 2_L} \otimes{\bf 3^0} \oplus{\bf 2_R} \otimes{\bf 3^0}. $$

The candidate is
\begin{itemize}
\item $\mathcal{A}_2 = \mathbb C \oplus \mathbb H \oplus M_3(\mathbb C) $
\item $\mathcal{H}_2 = M_F^{\oplus 3}, \qquad   M_F = \mathcal E \oplus \mathcal E^0$
\item $D_2 = \left(\begin{array}{cc}S & T^* \\T & \bar S\end{array}\right)$ (see reference for definition of $S$ and $T$)
\item $\gamma_2 =  c - J_2 c J_2,\quad c=(0,1,-1,0)\in\mathcal{A}_{LR}$ (defined by the $\mathbb Z_2-$grading of $\mathcal{H}$)
\item $J_2 (\xi, \bar \eta) =  ( \bar\xi, \eta),\quad \forall\xi,\eta\in\mathcal E$\\
\end{itemize}

Then we have the following results (among others) :
\begin{itemize}
\item Up to a finite abelian group, the group $SU(\mathcal{A}_2)$ is of the form
$$SU(\mathcal{A}_2) \sim U(1) \times SU(2) \times SU(3)$$
\item  The adjoint action of the $U(1)$ factor is given by multiplication of the basis vectors by the following powers of $\lambda\in U(1)$ :
$$\begin{array}{ccccc}
&\uparrow \otimes {\bf 1^0}  &\downarrow \otimes {\bf 1^0}  &\uparrow \otimes {\bf 3^0}  &\downarrow \otimes {\bf 3^0}\\
{\bf 2_L} & -1 & -1 & \frac 13 & \frac 13\\
{\bf 2_R} & 0 & -2 & \frac 43 & -\frac 23\\
\end{array}$$
\end{itemize}

\subsection*{The final Spectral Action}

The spectral triple product is done 
$$(\mathcal{A},\mathcal{H},D,\gamma,J) = (\mathcal{A}_1,\mathcal{H}_1,D_1,\gamma_1,J_1) \otimes(\mathcal{A}_2,\mathcal{H}_2,D_2,\gamma_2,J_2)$$
and thus is the final spectral action
$$S =\tr\left({ f\left({\frac{D}{\Lambda}}\right)}\right) + \left<{\psi, D \psi}\right>$$
where $\Lambda$ is the cut off parameter, $f$ a positive suitable function and the last term being added to handle the fermionic part.\\

So this action presents in the same time the bosonic and fermionic part of the Standard Model, minimally coupled to Euclidian Gravity.


\section[The Lorentzian Wall]{The Lorentzian Wall}

The most part of the current Connes' model is based on a Riemannian manifold describing Euclidian Gravity.  But of course this does not correspond  to any physical reality !  In order to build a physical version of the theory and apply it to real physical problems, one must find a Lorentzian version of the theory, and especially resolving the following two important problems : 

\begin{itemize}
\item The Hilbert space $\mathcal{H} = L^2(S)$ is constructed with an inner product
$$(\psi,\phi) = \int_M \psi^*\phi\, d\mu_g$$
where $d\mu_g$ is the measure associated to the metric $g$  on $M$. Such inner product becomes undefined (not positive definite) in case of  Lorentzian metric.
\item The distance formula
$$ d(x,y) = \sup_{f\in\mathcal{A}}\left\{{\left\vert{f(x) - f(y)}\right\vert : \left\Vert {[D,f]}\right\Vert  \leq 1}\right\}$$
is clearly Riemannian and cannot be used without knowledge of the causal structure of space-time.\\
\end{itemize}

This is what we could call the Lorentzian Wall : the theory is currently in a state that allows to study Euclidian models and make some first mathematical calculations, but from now the physic reality of such models or calculations seems to be out of sight, behind a not broken wall.\\

However, there exist at this time few attempts to generalize some particular elements of spectral triples to the Lorentzian case. These attempts are trying to make the first few breaks in the wall, but do not provide a complete solution.\\

These attempts are generally based on the hypothesis of a globally hyperbolic space-time.  This fact is comprehensible because there exists at this time no way to recover a Lorentzian metric from a spectral triple, so there is no accessible information on local causal structure of space-time, and a hypothesis on global causal structure is needed. \\

\subsection*{Spectral Quadruples}

[T. Kopf and M. Paschke, {\it A Spectral Quadruple for de Sitter Space}, J. Math. Phys. 43  (2002) 818-846]\\

The main idea of the spectral quadruple is to use the foliation of a globally hyperbolic space-time to define a family of commutative algebras
$$ \mathcal{A}_t = C^\infty(\Sigma_t)$$
acting by pointwise multiplication on a family of Hilbert spaces
$$\mathcal{H}_t = L^2(S_t)\qquad (a_t\psi)(x) = a_t(x)\psi(x)\quad\forall a_t\in \mathcal{A}_t, \psi \in \mathcal{H}_t, x \in\Sigma_t$$
where $S_t$ is the restriction of the $d+1$ dimensional spinor bundle over $\Sigma_t$.\\

The Dirac equation can be written in Hamiltonian form, using ADM formalism,
$$ H \psi = i \partial_t \psi$$
$$ H = -i N \left({ \omega_0^s + \gamma_c \gamma_b e_0^a e_i^b g^{ij} D_j^s - m\gamma_a e_0^a }\right) + N^i\partial_i.$$

The quadruple needs to add as data a time-direction operator $E$,  and $H$ allows to define a time-evolution operator $\mathcal{H}_{t_1} \rightarrow \mathcal{H}_{t_2}$.\\ 

But therefore, there are clearly different Hamiltonians which describe the same space-time.  This problem is (tried to be) solved by considering not one given, but all Hamiltonians, and then by remarking that - if we consider algebras $\mathcal{A}_t$ as objects of a category - time evolution operator for each Hamiltonian form a invertible morphism, so they form together a groupoid.\\

So a spectral quadruple is the data of
\begin{itemize}
\item a family of algebras $A_t$ 
\item an Hilbert space $\mathcal{H}$ 
\item a groupoid $G$
\item a charge conjugation operator $C$
\item an element $\gamma(t)$
\item a time vector $E(t)$
\end{itemize}

Current problems of this attempt :
\begin{itemize}
\item If the groupoid is very small ($G \cong \mathbb R$), then there would be only one Hamiltonian and only one time direction, so all data at time 0 are sufficient to determine the spectral quadruple, and there is a real problem of missing covariance.  However, if $G$ is chosen larger (all possible time evolutions in the extreme case), more algebras $\mathcal{A}_t$ are needed in the data, and inputs become very consequent.
\item It is not proved that all Hamiltonians derived from $G$ describe the same manifold.
\item There is not at this time a total Dirac operator (just a spacial one can be derived from the quadruple) and so no way to reconstruct the metric.\\
\end{itemize}

\subsection*{Lorentzian distance}

[V. Moretti, {\it Aspects of noncommutative Lorentzian geometry for globally hyperbolic spacetimes}, Rev. Math. Phys. 15 (2003) 1171-1217]\\

This is a proposition of a generalization of the distance formula
$$ d(x,y) = \sup_{f\in\mathcal{A}}\left\{{\left\vert{f(x) - f(y)}\right\vert : \left\Vert {[D,f]}\right\Vert  \leq 1}\right\}$$
to the Lorentzian case.\\

The main problem to generalize this formula is the absence of a well definite metric (so no more Lipschitz condition, which plays a crucial role in the proof of the formula in the Euclidian case). The lack of information on causal structure of spacetime and the characteristics of Lorentzian distances are also both problematic.\\

The distance proposed is the following, in a globally hyperbolic space-time and for $q$ in the causal future of $p$ :
$$  d(p,q) = \inf\left\{{\left\vert{f(q) - f(p)}\right\vert : f \in \mathcal C_{\mu_g}(\bar I), I \in \mathcal X,\ p,q \in \bar I, \left\Vert {[f,[f,\frac{\triangle}{2}]^{-1}]}\right\Vert _I \leq 1}\right\}$$
where $ \mathcal C_{\mu_g}(N)$ denotes the class of causal functions on $N \subset M$ (i.e. continuous map which does not decrease along every causal future-directed curve contained in $N$) which are smooth almost everywhere in $N$, and $\mathcal X$ denotes the class of all regions $I\in M$ which are open, causally convex such that $\bar I$ is compact and with 0-measure boundary. A Laplace-Beltrami operator is used instead of a Dirac one.\\

Current problems of this attempt :
\begin{itemize}
\item The minimality of the axioms is not verified.
\item There is at this time no concrete non-commutative models to illustrate the formalism.
\item The problems of the Dirac operator and the inner product are not solved.
\item Information on causal structure is only contained in the set of causal functions $ \mathcal C_{\mu_g}(N)$ which seems complicated to build.\\
\end{itemize}

But despite this incompleteness, Moretti's study provides us a quite complete and interesting review of the crucial problem of the knowledge of local causal structure in the algebraization of Lorentzian geometry.\\

\subsection*{Krein Spaces}

[A. Strohmaier, {\it  On noncommutative and pseudo-Riemannian geometry}, J. Geom. Phys. 56 (2006) 175-19]\\

The main idea is to define a fundamental symmetry on spectral triples which will have a similar effect than a Wick rotation to a metric.\\

We said before that the inner product
$$(\psi,\phi) = \int_M \psi^*\phi\, d\mu_g$$
is not positive definite, but the space with some hypotheses can define a Krein space.\\

A indefinite inner product on $V$ is a map $V \times V \rightarrow \mathbb C$ which satisfies 
$$(v,\lambda w_1 + \mu w_2) = \lambda (v,w_1) + \mu(v, w_2), \qquad \overline{(v,w)} = (w,v).$$
It is said non-degenerated if
$$(v,w)=0 \quad\forall v\in V \ \ \Rightarrow\ \  w = 0.$$

If $V$ can be written as the direct sum of orthogonal spaces $V^+$ and $V^-$ such that the inner product is positive definite on $V^+$ and negative definite on $V^-$ and the two subspaces are complete in the norms induced, it is called a Krein space.\\

In such case, the operator $\mathcal J = \text{id} \oplus -\text{id}$ defines a positive definite inner product by
$$\left<{\cdot,\cdot}\right>_{\mathcal J} = (\cdot,\mathcal J\cdot).$$
This is our fundamental symmetry. So it is possible to redefine a spectral triple based on a Krein space and with a Dirac operator no more self-adjoint but {\it Krein-selfadjoint} (generalization of selfadjointness  in Krein spaces).\\

Current problems of this attempt :
\begin{itemize}
\item Because the Dirac operator is not elliptic as in the Riemannian case, singularities can appear in the differential structure.
\item Although differentials and integrals are conserved, there is no Lorentzian equivalent distance formula.
\item It is not know if the Einstein-Hilbert action can be recovered in this formalism.
\item It is not know if there exist additional conditions which can guarantee that the final action is independent of the choice of $\mathcal J$.
\item The spectral triples are based on compact pseudo-Riemannian manifolds, or compact Lorentzian manifolds should clearly be avoided (but compactness could be only a technical problem).\\
\end{itemize}

It is clear that Krein spaces do not provide us a way to transcribe the complete theory in a not positive definite framework. We need concrete elements replacing the distance formula and the Dirac operator. Nevertheless, this is a well defined and successful manner to deal with the not positive definite problem, so the idea of Krein space could be an essential element in the construction of a complete Lorentzian theory.


\section[Conclusions]{Conclusions}

Connes' theory of Non-Commutative Geometry is a well defined mathematical tool used to describe in a unified algebraic framework both geometric theories and gauge theories.\\

Significant results have be done, especially in the way of approaching Standard Model of particles coupled with Riemaniann geometry.\\

But although there are really good advances around the Standard Model, the gravitational part is the poor student, especially because a Lorentzian version of the theory is still out of sight.\\

In particular, it should be interesting to solve these problems :
\begin{itemize}
\item Find a completely well defined equivalent of the Hilbert space in case of missing positive definite inner product.
\item Solve the problem of missing information on locally causal structure in the algebraic framework.
\item Propose a manner to extract information on the metric or distance from a Dirac operator or equivalent, manner which must correspond to the usual Lorentzian metric in the geometrical case.\\
\end{itemize}

{\bf General book references :}
\begin{itemize}
\item A. Connes, {\it  Noncommutative Geometry}, Academic Press (1994)
\item J.M. Garcia-Bondia, J.C. Varilly and H. Figueroa, {\it Elements of non-commutative Geometry}, Birkhauser (2001)
\item G. Landi, {\it An Introduction to Noncommutative Spaces and their Geometries}, Lecture Notes in Physics, Spinger (1997)
\end{itemize}

\end{document}